\documentclass[conference]{IEEEtran}
\IEEEoverridecommandlockouts
\usepackage{cite}
\usepackage{amsmath,amssymb,amsfonts}
\usepackage{algorithmic}
\usepackage{graphicx}
\usepackage{textcomp}
\usepackage{xcolor}

\usepackage{subfigure}
\usepackage{url}
\usepackage{paralist}
\usepackage{enumitem}
\usepackage{xspace}
\usepackage{amsthm, amssymb}
\usepackage{algorithm}

\usepackage{wrapfig}
\usepackage{tabularray}

\usepackage{orcidlink}

\newcommand{\ourTool}{\emph{DEM}}
\newcommand{\mysubsection}[1]{\medskip\noindent\textbf{#1}}
\newcommand{\plr}{\text{plr}\xspace{}}
\DeclareMathOperator*{\argmmax}{arg\,max}

\theoremstyle{definition}
\newtheorem{definition}{Definition}[section]

\def\BibTeX{{\rm B\kern-.05em{\sc i\kern-.025em b}\kern-.08em
    T\kern-.1667em\lower.7ex\hbox{E}\kern-.125emX}}
\begin{document}

\title{DEM: A Method for Certifying  Deep Neural Network
	Classifier Outputs in Aerospace\\
        \thanks{This work was partially funded by the European Union
          (ERC, VeriDeL, 101112713). Views and opinions expressed are
          however those of the author(s) only and do not necessarily
          reflect those of the European Union or the European Research
          Council Executive Agency. Neither the European Union nor the
          granting authority can be held responsible for them.  The
          work was further 
          partially funded by an NSFC-ISF grant to David Harel, issued jointly
          by the National Natural Science Foundation of China (NSFC)
          and the Israel Science Foundation (ISF grant 3698/21), and
          in part by the Minerva grant 714132. Additional support was
          provided by a research grant to DH from Louis J. Lavigne and
          Nancy Rothman, the Carter Chapman Shreve Family Foundation,
          Dr. and Mrs. Donald Rivin, and the Estate of Smigel Trust.
        }
     }

\author{\IEEEauthorblockN{Guy Katz}
\IEEEauthorblockA{\textit{The Hebrew University of Jerusalem } \\
Jerusalem, Israely \\
gkatz@mail.huji.ac.il}
\and
\IEEEauthorblockN{Natan Levy}
\IEEEauthorblockA{\textit{The Hebrew University of Jerusalem } \\
Jerusalem, Israely \\
natan.levy1@mail.huji.ac.il}
\and
\IEEEauthorblockN{Idan Refaeli}
\IEEEauthorblockA{\textit{The Hebrew University of Jerusalem } \\
Jerusalem, Israely \\
idan.refaeli@mail.huji.ac.il}
\and
\IEEEauthorblockN{Raz Yerushalmi \orcidlink{0000-0002-0513-3211}}
\IEEEauthorblockA{\textit{The Weizmann Institute of Science} \\
Rehovot, Israel \\
\textit{The Hebrew University of Jerusalem } \\
Jerusalem, Israely \\
raz.yerushalmi@weizmann.ac.il}

}

\maketitle

\begin{abstract}
	Air transportation, a critical component of modern life, faces significant challenges concerning efficiency, environmental sustainability, and safety. Addressing these challenges necessitates innovative solutions, such as using deep neural networks (DNNs). However, although DNNs demonstrate remarkable performance, they remain susceptible to tiny perturbations in their inputs, which may result in misclassification.
	The vulnerability, known as adversarial inputs, may trigger a chain of events that could result in significant, or even catastrophic, failures.
	In this regard, adversarial inputs constitute a crucial obstacle to the integration of DNNs into safety-critical aerospace systems.
	
	In the present work, we introduce a novel, output-centric
        method for certifying DNNs that addresses this challenge. This
        method utilizes statistical techniques to flag out specific
        inputs for which the DNN's output might be unreliable
        so that a human expert may examine them. In contrast to existing techniques, which typically attempt to certify the entire DNN, the proposed method certifies specific outputs. Moreover, the proposed method uses the DNN as a black box and makes no assumptions about its topology.

	We developed a proof-of-concept tool called \ourTool ~to demonstrate the feasibility of the proposed method.
	For evaluation, we tested the proposed method on a VGG-16 model, trained on the CIFAR16 dataset, showing that the proposed method achieved a high percentage of success in detecting adversarial inputs.
	We believe this work constitutes another step towards integrating deep neural networks in safety-critical applications --- especially in the aerospace domain, where high standards of quality and reliability are crucial.
\end{abstract}


\section{Introduction}
Air transportation handles billions of passengers annually, and
requires advanced solutions for addressing long-standing efficiency,
environmental impact, and safety challenges~\cite{EuUnAvSaAg23}. The
deep learning revolution shows significant potential for addressing
these requirements, e.g., through the use of deep neural networks
(DNNs)~\cite{PaDaJo03,XuChHeZhDu19}. However, although DNNs have
demonstrated remarkable performance, they remain susceptible to tiny
perturbations in their inputs, which might result in
misclassification.  This vulnerability, referred to as
\emph{adversarial inputs}~\cite{GoShSz15,KaBaDiJuKo17}, could trigger
a chain of events that might result in serious injury, extensive
damage to valuable assets, or irreparable damage to delicate
ecosystems~\cite{Kn02}. Consequently, adversarial inputs pose a
significant barrier to the integration of DNNs in a variety of fields,
such as safety critical systems.

In the aerospace industry, certification authorities such as the
European Aviation Safety Agency (EASA) and the Federal Aviation
Administration (FAA) play a key role in advocating civil aviation
safety and environmental preservation. These authoritative bodies
acknowledge the immense value of incorporating DNNs into
safety-critical applications~\cite{EuUnAvSaAg23}, yet also emphasize
that conventional software certification guidelines, such as
DO-178~\cite{FAA13}, are not applicable to DNNs. In recognition of the
inherent challenges of certifying DNNs within the aerospace domain,
EASA recently released its comprehensive AI
roadmap~\cite{EuUnAvSaAg23}, which outlines seven pivotal requirements
for ensuring artificial intelligence trustworthiness. Of these,
\emph{robustness to adversarial conditions} is mentioned as one of the
fundamental prerequisites.  Thus, it is evident that ensuring a DNN's
robustness to adversarial inputs will play a crucial role in any
future certification process.

During the preliminary stage of an aerospace certification process,
the SAE ARP-4754 guidelines~\cite{FAA10} call for conducting a
\emph{safety analysis} procedure for all aircraft systems and
subsystems. For each subsystem, the safety engineering team must
determine the permissible probability of failure, with which the
design engineer is required to comply. In order to carry out such a process for a system that incorporates a
DNN, the engineering team must demonstrate that the probability of
a failure related to the DNN  does not exceed the threshold established in the
safety analysis. Unfortunately, there is currently a shortage of
techniques for performing this analysis that are both sufficiently
precise and scalable.

Here, we seek to bridge this crucial gap by introducing a novel DNN
certification approach, \emph{DNN Enable Monitor} (\ourTool).
Inspired by wireless network management
concepts~\cite{TsTsHsYa18,LeMaShEl22,AsKa23}, instead of certifying
the DNN as a whole, \ourTool{} focuses on certifying the concrete
DNN's outputs. For each DNN output, \ourTool{} calculates during
inference time the probability of misclassification --- and if this
probability is below a certain acceptable threshold, the output is
considered reliable. Otherwise, the output is flagged and passed on
for further analysis, e.g., by a human expert. The motivation here is
that well-trained DNNs, working under normal conditions, are generally
correct; and so, identifying the few cases where the DNN might be
wrong, and further analyzing only these cases, can relieve a
significant cognitive load off the human experts, compared to a manual
approach.  Further, if the automated certification is accurate enough,
the entire process becomes nearly fully automatic --- and also
sufficiently accurate to meet the bars specified in the safety
analysis phase. The proposed method is in line with other
semi-automatic safety analysis and optimization methods for
safety-critical applications~\cite{MuAbThNoAmRa18}.

In order to effectively distinguish between regular input and an
adversarial input, \ourTool{} leverages the concept of
\emph{probabilistic global categorial robustness}, presented as part
of the \emph{gRoMA} method for statistically evaluating the global
robustness of a classifier DNN, per output category~\cite{LeYeKa23}.
To gain intuition, consider an input $\vec{x_0}$, which is mapped by
the DNN to label $l_0$. If this classification is correct, most input
points in the robustness region around $\vec{x_0}$ would also be
classified as $l_0$~\cite{LeYeKa23}. If, however, $\vec{x_0}$ is an
adversarial input, which can intuitively be interpreted as a
``glitch'' in the DNN, then other points around $\vec{x_0}$ would be
classified differently. \ourTool{} is used to effectively distinguish
between these two cases.

For evaluation purposes, we created a proof-of-concept implementation
of \ourTool, and used it to evaluate VGG16~\cite{SiZi15} and
Resnet~\cite{HeZhReSu16} DNN models. Our evaluation shows that
\ourTool{} outperforms state-of-the-art methods in adversarial input
detection, and that it successfully distinguishes adversarial inputs
from genuine inputs with a very high success rate in some
categories. For these categories, success rates are sufficiently high
to meet the aerospace regulatory requirements.

A major advantage of \ourTool{} is that it computes different
thresholds for certifying outputs for the different classifier
categories --- making it flexible and accurate enough to handle cases
where it is impossible to select a uniform threshold for all output
categories.  To the best of our knowledge, this is the first effort at
certifying the categorial robustness of DNN outputs, for
safety-critical DNNs.

The rest of the paper is organized as follows: We begin in
Sec.~\ref{RelatedWrok} with a description of prior work, and then
present the required background on adversarial robustness in
Sec.~\ref{Background}. In Sec.~\ref{Proposed-Method} we describe our proposed method for
measuring adversarial robustness,
followed by a description of our evaluation in
Sec.~\ref{Evaluation}. Finally, in Sec.~\ref{Discussion}, we summarize and discuss our results.

\section{Related Work}
\label{RelatedWrok}
The certification of DNNs and their integration into safety-critical
applications has been the subject of extensive
research~\cite{GaShTiWi23,DmScHo23,DmScHo21}. 
In general, DNN certification rests on two main foundations: certification guidelines
and DNN robustness.

\medskip
\noindent
\textbf{Certification Guidelines.} 
A widely accepted cornerstone in
system and software certification within the aerospace domain are the
``ARP-4754 --- Guidelines for Development of Civil Aircraft Systems
and Equipment Certification'' guidelines~\cite{FAA10}. These
guidelines were authored by SAE International, a global association of
aerospace engineers and technical experts. The guidelines focus on
safety considerations when developing civil aircraft and their
associated systems. However, these guidelines do not directly apply to
DNN components.

A recent work~\cite{HuHuHuPe21} proposed an approach to assessing DNN
robustness through an alternative Functional Hazard Analysis (FHA)
method. This method seeks to show that the likelihood
that a neural network performs unexpectedly is below an
acceptable threshold, defined by the user, with a confidence level of 99\%.
\ourTool, on the other hand, measures the probability of a failure condition,
given a specific input. \ourTool{} meets the probabilistic goals of the ARP-4754 guidelines
objectives, and could potentially be used to certify DNN-based components
without further adjustments.

Another recent study~\cite{GaShTiWi23} developed a comprehensive framework of
principles for DNN certification, which is aligned with ARP-4754 ---
but which does not include a concrete method or tool that can be used.
While ARP-4754 focuses on system level certification, the influential
DO-178 guidelines for airborne systems focus on software
certification~\cite{FAA13}. The DO-178 guidelines supplement ARP-4754's safety requirements with 5 levels (A-E) for
software failure conditions. Initial work has discussed certifying
DNNs to level D \cite{DmScHo21,GaShHo22} or C
\cite{DmScBoAbHo23,DmScHo23}, 
but to date none have reached Level A --- the most severe level, required where failures may
lead to fatalities, which is the case in aerospace. We hope that our
approach can be used to satisfy Level A requirements in the
future.

\medskip
\noindent
\textbf{DNN Robustness.} 
An alternative approach to the one proposed
here, which has received significant attention, is the formal
certification of DNN robustness. The idea is to conclude, a-priori,
that a DNN is safe to use, at least for certain regions of its input
space.

One approach for drawing such conclusions is to \emph{formally verify}  DNNs~\cite{KaBaDiJuKo17,WaPeWHYaJa2018}, i.e., mathematically prove that the DNN behavior
adheres to a given specifications. DNN verification
methods are highly accurate sound and often complete~\cite{AmWuBaKa21}, but suffer from limited scalability,
require white-box access to the DNN in question, and pose certain
limitations on the DNN's topology and activation functions. 

Another approach attempts to circumvent these limitations by
certifying a DNN's robustness with significant margins of
error~\cite{ZhWeJo2022,WeChNgSqBoOsSa19,AnSo20,HuHuHuPe21}. 
This kind of attempt may be inadequate for aerospace certification, where large error margins are usually not acceptable.

Finally, there exist statistical approaches for evaluating the
probability that a specific input is being misclassified by the
network (i.e., is an adversarial input)~\cite{WeRaTeYe18,TiFuRo2021,MaNoOr19,CoRoZi19,MaLiWaErWiScSoHoBa18}.  
\ourTool{} is statistical at its core as well. It can be configured to be more conservative, i.e., to prefer type II error  over type I error, or less, depending on the use case. 
A detailed comparison between \ourTool{} and a state-of-the-art technique~\cite{MaLiWaErWiScSoHoBa18} is given in section~\ref{Evaluation}.

\section{Background}

\label{Background}
\mysubsection{Deep Neural Networks (DNNs).}
A deep neural network (DNN)
$N$ is a function
$N: \mathbb{R}^n \rightarrow \mathbb{R}^m$, which maps an input vector
$ \vec{x} \in \mathbb{R}^n$ to an output vector
$\vec{y} \in \mathbb{R}^m$. In this paper we focus on classification
DNNs, where $\vec{x}$ is classified as class $c$ if the $c$'th entry
of $N(\vec{x})$ has the highest score:
$\argmmax(N(\vec{x}))=c$.

\mysubsection{Local Adversarial Robustness.} 
Local adversarial
robustness is a measure of how resilient a DNN is to perturbations around
specific input points~\cite{BaIoLaVyNoCr16}:

\begin{definition}
	\label{definition1}
	A DNN $N$ is $\epsilon$-locally-robust at input point $\vec{x_0}$ iff
	\[
	\forall \vec{x}.
	\displaystyle || \vec{x} -\vec{x_0} ||_{\infty} \le \epsilon 
	\Rightarrow \argmmax(N(\vec{x})) = \argmmax(N(\vec{x_0})) 
	\]
\end{definition}
Intuitively, Definition~\ref{definition1} states that for input vector
$\vec{x}$, the network assigns to $\vec{x}$ the same label that it assigns
to $\vec{x_0}$, as long as the distance of $\vec{x_0}$ from $\vec{x}$ is at most $\epsilon$ (using the $L_\infty$ norm). 

\mysubsection{Probabilistic Local Robustness.}
Definition~\ref{definition1} is Boolean: given $\epsilon$ and
$\vec{x_0}$, the DNN is either robust or not robust. However, in
real-world settings, and specifically in aerospace applications, systems
could still be determined to be sufficiently robust if the likelihood
of encountering adversarial inputs is greater than zero, but is
sufficiently low.  Federal agencies, for example, provide guidance
that a likelihood that does not exceed $10^{-9}$ (per
operational hour under normal conditions) for an extremely improbable failure conditions event is
acceptable~\cite{LaNi2011}.  Consequently, prior research suggested an adjustment to
Definition~\ref{definition1} that allows to reason about a DNN's
robustness in terms of its probabilistic-local-robustness
(\plr)~\cite{LeKa2021}, which is a real value that indicates a DNN's
resilience to adversarial perturbations imposed on specific
inputs. More formally:

\begin{definition}
	\label{definition2}
	The $\epsilon$-probabilistic-local-robustness (PLR) score of a DNN $N$ at input point $\vec{x_0}$, abbreviated $\plr{}_{\epsilon}(N,\vec{x_0})$, is defined as:
	\begin{align*}
		\plr{}_{\epsilon}(N,\vec{x_0})
		&\triangleq 
		P_{\vec{x}:  \lVert \vec{x} - \vec{x_0} \rVert_\infty \le \epsilon} \\
		&\quad \left[
		\arg\max(N(\vec{x})) = \arg\max(N(\vec{x_0}))
		\right]
	\end{align*}
\end{definition}

Intuitively, the definition measures the probability that for a specific input
$\vec{x_0}$, an input $\vec{x}$ drawn at random from the
$\epsilon$-region around $\vec{x_0}$ will have the same label as
$\vec{x_0}$.

\mysubsection{Probabilistic Global Caterorial Robustness (PGCR).}
Although Definition~\ref{definition2} is more realistic than
Definition~\ref{definition1}, it suffers from two drawbacks. The
first: recent studies indicate a significant disparity in robustness
among the output categories of a DNN~\cite{LeKa2021,LeYeKa23}, which
is extremely relevant to safety-critical applications.
However, Definition~\ref{definition2}
does not distinguish between output classes. The second drawback is
that Definition~\ref{definition2} considers only local robustness,
i.e., robustness around a single input point, in a potentially vast
input space; whereas it may be more realistic to evaluate robustness
on large, continuous chunks of the input space. To address these
drawbacks, we use the following definition~\cite{LeYeKa23}:

\begin{definition}
	\label{definition3}
	Let $N$ be a DNN, let $l \in L$ be an output label. 
	The ($\epsilon,\delta$)-PGCR score for $N$ with respect to $l$, denoted
	$pgcr_{\delta,\epsilon}(N,l)$, is defined as:
	\begin{align*}
		pgcr_{\delta,\epsilon}(N,l) &\triangleq  
		P_{\vec{x_1}, \vec{x_2} \in \mathbb{R}^n, \lVert \vec{x_1} - \vec{x_2} \rVert_{\infty} \le \epsilon} \\
		&\quad \left[
		| N(\vec{x_1})[l] - N(\vec{x_2})[l] | < \delta 
		\right]
	\end{align*}
\end{definition}

This definition captures the probability that for an input
$\vec{x_1}$, and for an input $\vec{x_2}$ that is at most
$\epsilon$ apart from $\vec{x_1}$, the confidence scores for
inputs $\vec{x_1}$ and $\vec{x_2}$ will differ by at most
$\delta$ for the label $l$.

\mysubsection{Hypothesis Testing.}  Hypothesis testing is the task of
accepting the \emph{null hypothesis} $H_{0}$, or rejecting it in favor
of an alternative hypothesis $H_{1}$. To achieve this, we employ a
statistical test $h$, which is a function of the measurements. We further 
select a threshold $\tau$, such that $h > \tau$ implies that $H_{0}$
is rejected in favor of $H_{1}$, whereas $h \le \tau$ implies that
$H_{0}$ is accepted.

It is common to consider the \emph{significance} and \emph{power} of a
test in order to assess its usefulness:
\begin{itemize}
\item The significance $\alpha$ is given as
  $\alpha = P(h > \tau \mid H_{0})$, i.e.., the probability for a
  false positive, leading to the rejection of $H_{0}$ although it is
  in fact true  (type I error).
\item The power is given by $ 1 - \beta$, where
  $\beta = P(h \le \tau \mid H_{1})$ is the probability for a false
  negative, i.e., of wrongfully rejecting $H_{1}$ (type II error).
  The power can also be formulated directly as $P(h > \tau \mid H_{1})$,
  i.e., the probability for a true positive, or of correctly rejecting $H_{0}$, given that $H_{1}$ is true.
\end{itemize}

In general, we seek to reduce both $\alpha$ and $\beta$. Following the
Neyman-Pearson lemma~\cite{NePe28}, one may maximize the power
(minimize $\beta$) given $\alpha$:

\[
  h = \frac{\mathcal{L} ( H_{1} \mid \text{measurements} ) }{ \mathcal{L}(
    H_{0} \mid \text{measurements})}
\]
where $\mathcal{L} (H \mid \text{measurements} )$ is
the likelihood of $H$ given the measurements, and is equal to the conditional
probability $P(\text{measurements} \mid H)$.  In this work, we employ such
likelihood testing to decide between the hypotheses, to determine
whether a given input is adversarial or not.

\section{The Proposed Method}
\label{Proposed-Method}

\subsection{The Inference Phase}

Given an input point $\vec{x}_0$, our objective is to quantify the
reliability of the DNN's particular prediction, $N(\vec{x}_0)$; i.e.,
to accept or reject the $H_{0}$ hypothesis that the input is adversarial. Our
method for achieving this consists of the following steps:
\begin{itemize}
	\item Execute the DNN on input $\vec{x}_0$, and obtain the prediction $N(\vec{x}_0)$.
	\item Generate $k$ random perturbations around $\vec{x}_0$, denoted
	$\vec{x}'_1\ldots \vec{x}'_k$, sampled uniformly at random from an
	$\epsilon$-region around $\vec{x}_0$.
	\item For each sampled perturbation point $\vec{x}_i'$, compute the
	prediction $N(\vec{x}_i')$ and check whether it coincides with
	$N(\vec{x}_0)$. Count the number of such matches (``number of
	hits''), denoted as $h$.
	\item If $h$ is \emph{above} a certain threshold $\mathcal{T}$, certify the
	prediction of $N(\vec{x}_0)$ as correct, i.e., reject the $H_{0}$ hypothesis. Otherwise, flag it as suspicious.
\end{itemize}

The motivation for this approach is based on \ref{definition3}. Intuitively, for a genuine prediction over the input $\vec{x}_0$, the network would produce stable predictions in the $\epsilon$-region around $\vec{x}_0$; in such a case, the number of hits will be high and will exceed the threshold $\mathcal{T}$. 
Conversely, if $\vec{x}_0$'s prediction is not genuine, i.e.~$\vec{x}_0$ is an
adversarial input for $N$, then the network would produce different predictions in the $\epsilon$-region around $\vec{x}_0$, and the number of hits will be lower.

Naturally, in order for this approach to work, care must be taken
in determining the two key parameters: $\epsilon$, which determines the size
of the region around $\vec{x}_0$ from which perturbations are sampled,
and the threshold $\mathcal{T}$ for deciding  whether a prediction is 
correct. Below we present a method for selecting these values.

\subsection{Dataset Preparation and Calibration}
\label{DatasetPreParation}

\begin{figure*}[htb]
%\lipsum[1-2]
	\centering
	\subfigure[Airplane, Automotive, Bird, Cat, Deer categories, respectively]
	{\label{first_row}
		\includegraphics[trim = 0.cm 0.5cm 1cm 1cm,clip,width=.19\textwidth]{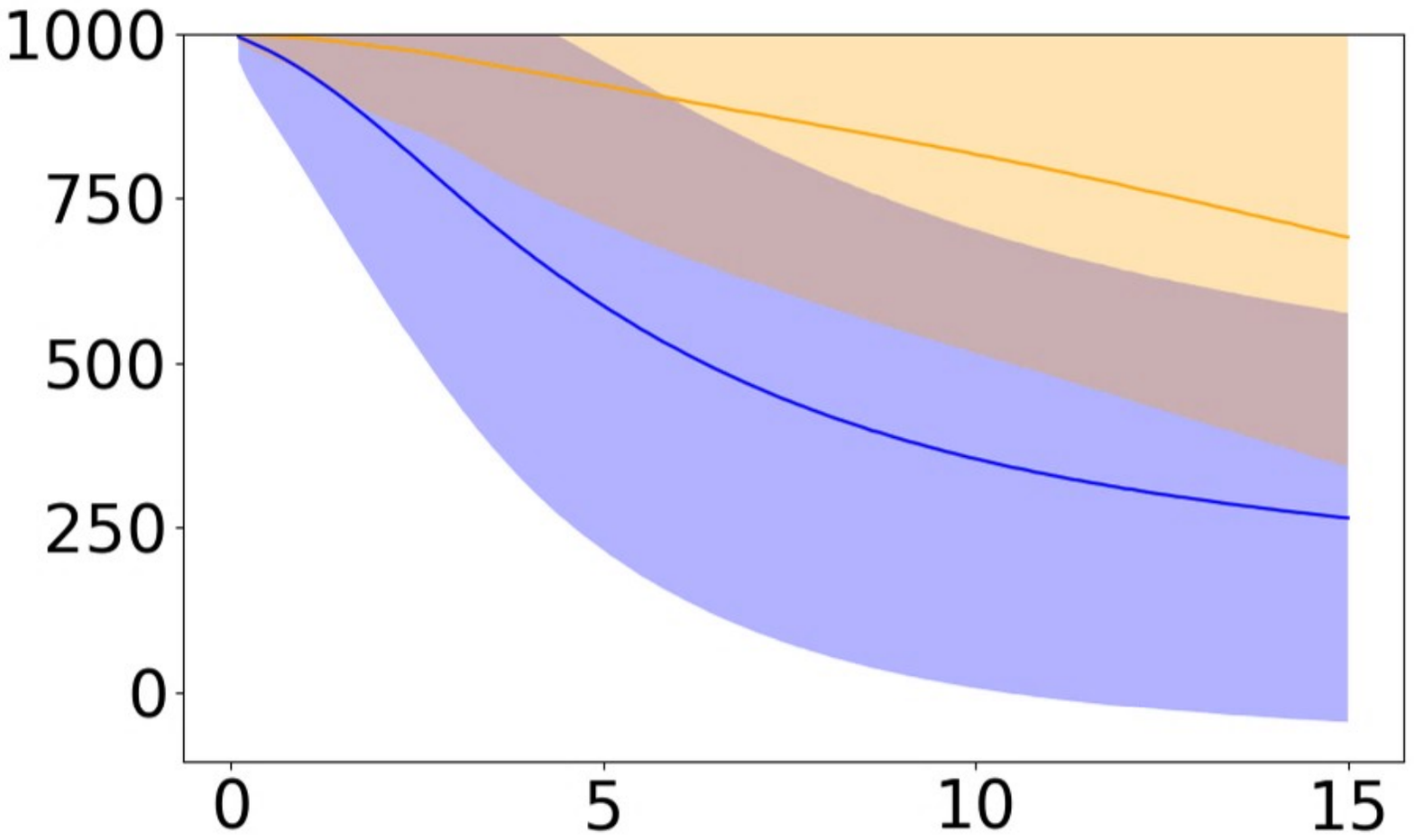}
		\includegraphics[trim = 0.5cm 0.5cm 1cm 1cm,clip,width=.19\textwidth]{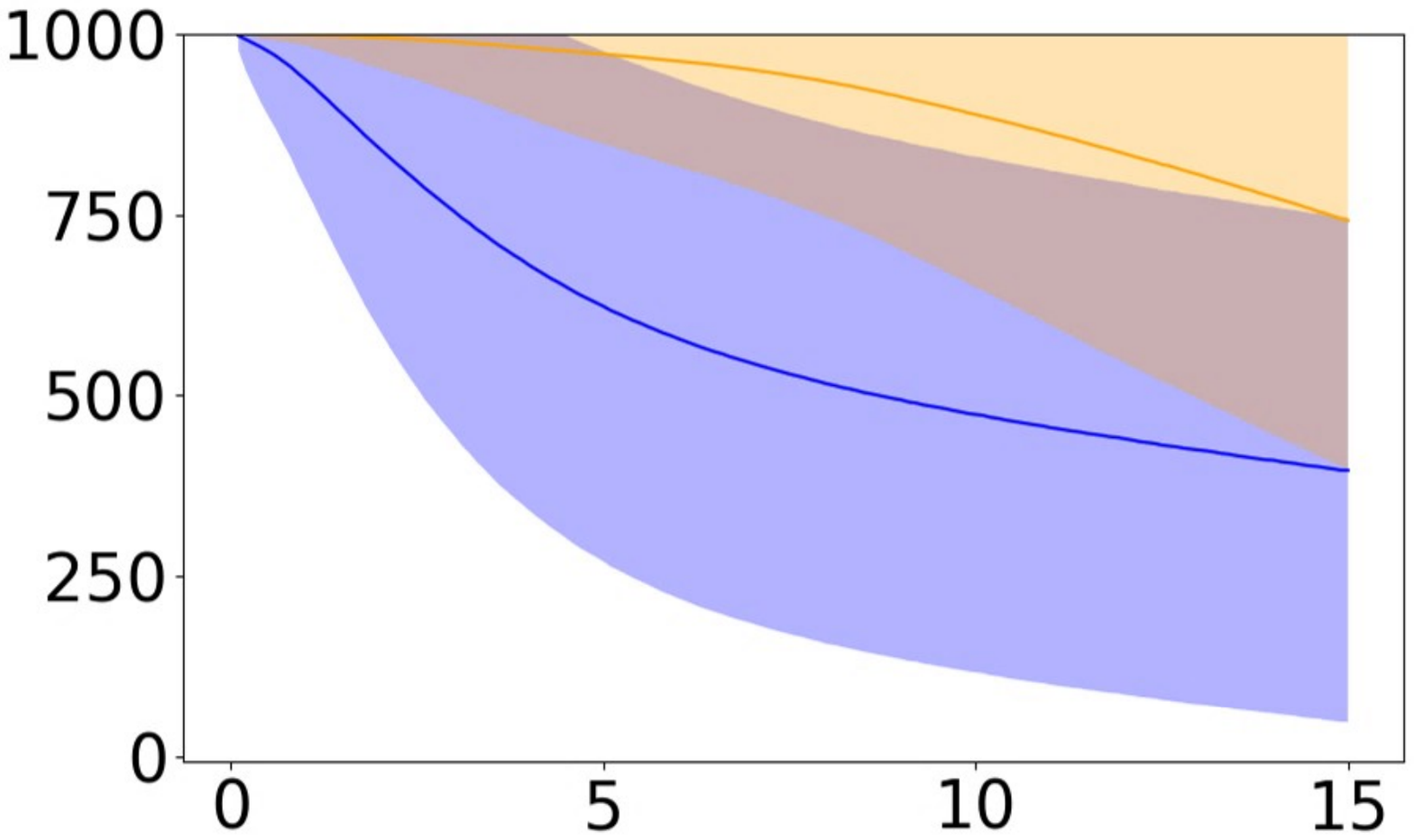}
		\includegraphics[trim = 0.5cm 0.5cm 1cm 1cm,clip,width=.19\textwidth]{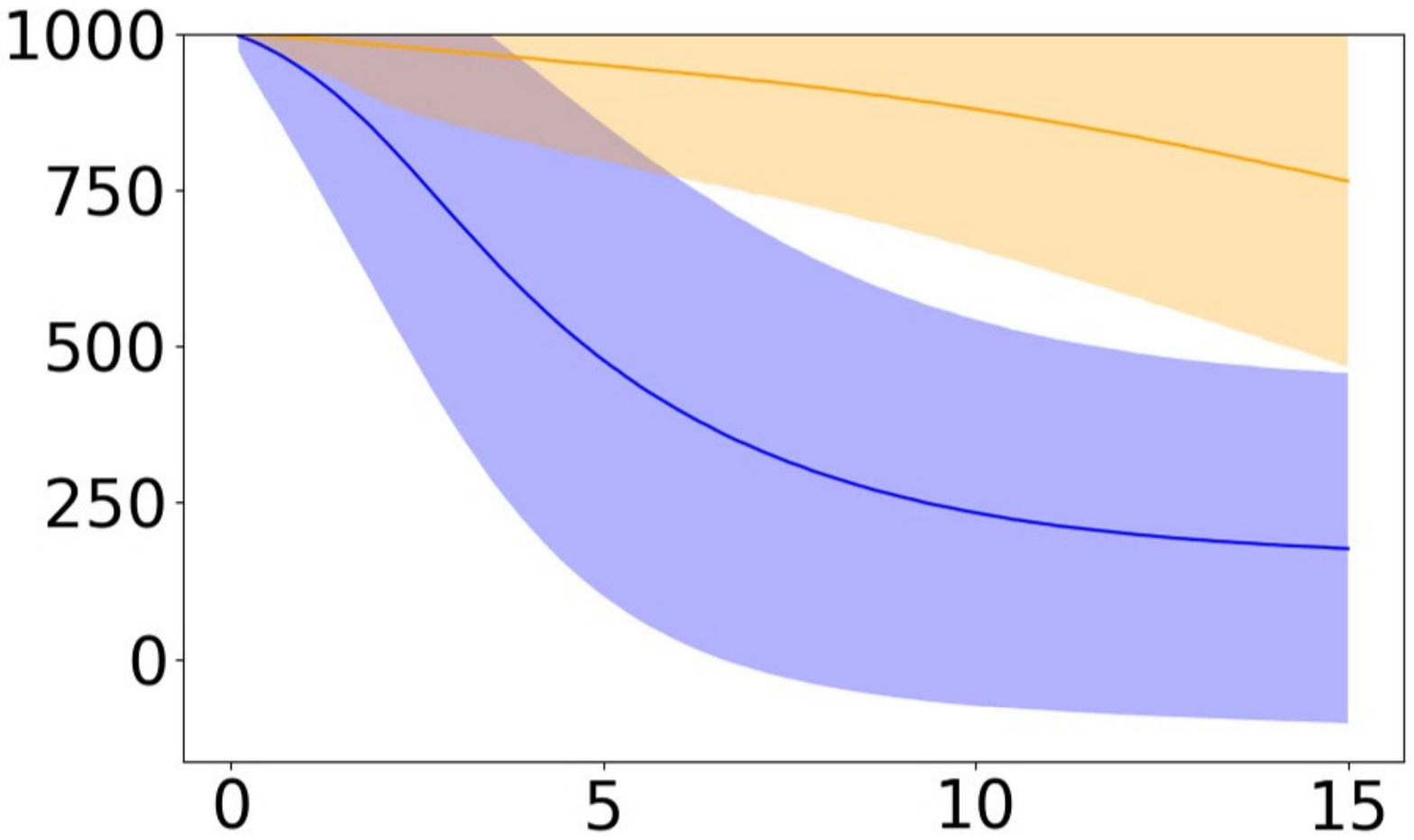}
		\includegraphics[trim = 0.5cm 0.5cm 1cm 1cm,clip,width=.19\textwidth]{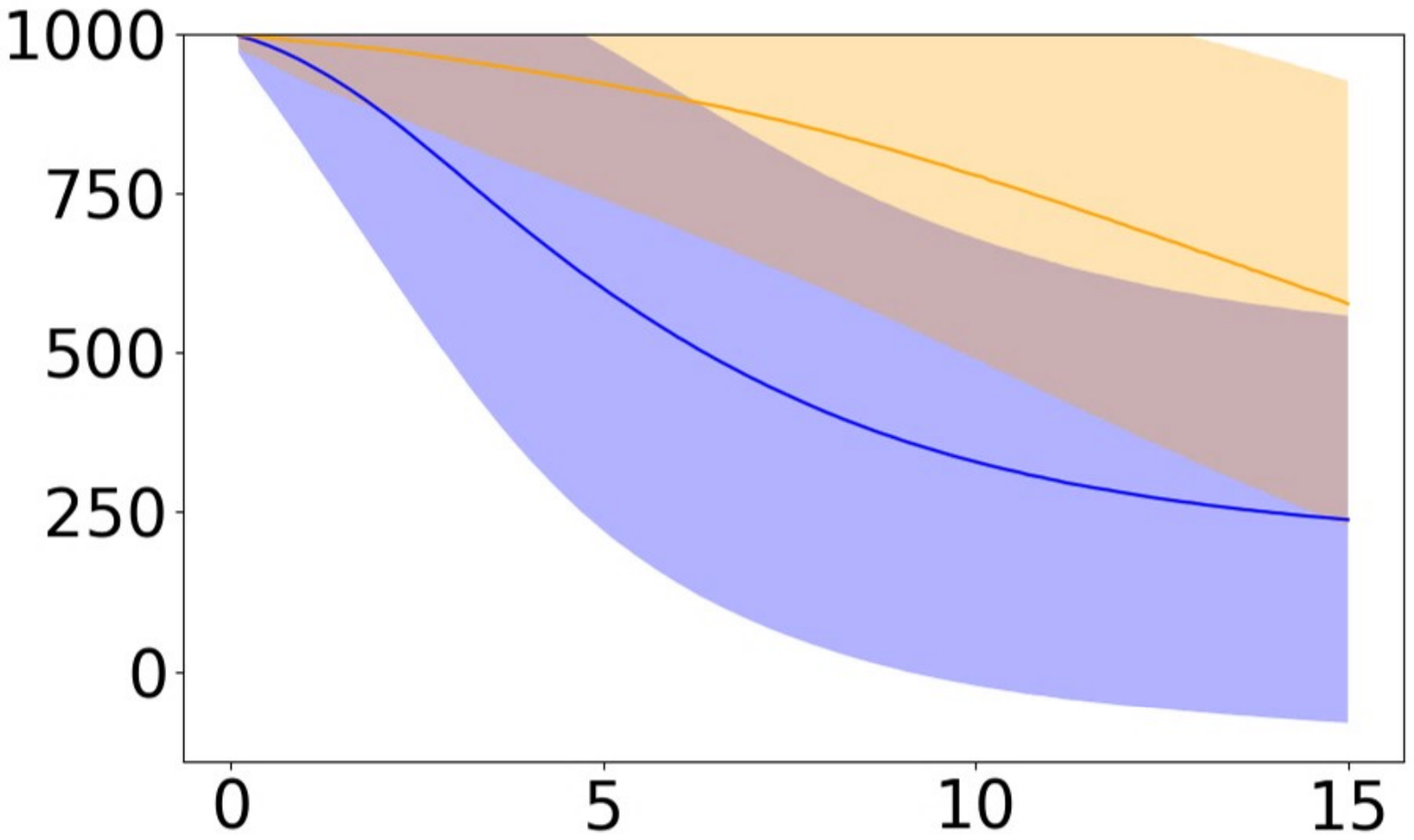}
		\includegraphics[trim = 0.5cm 0.5cm 1cm 1cm,clip,width=.19\textwidth]{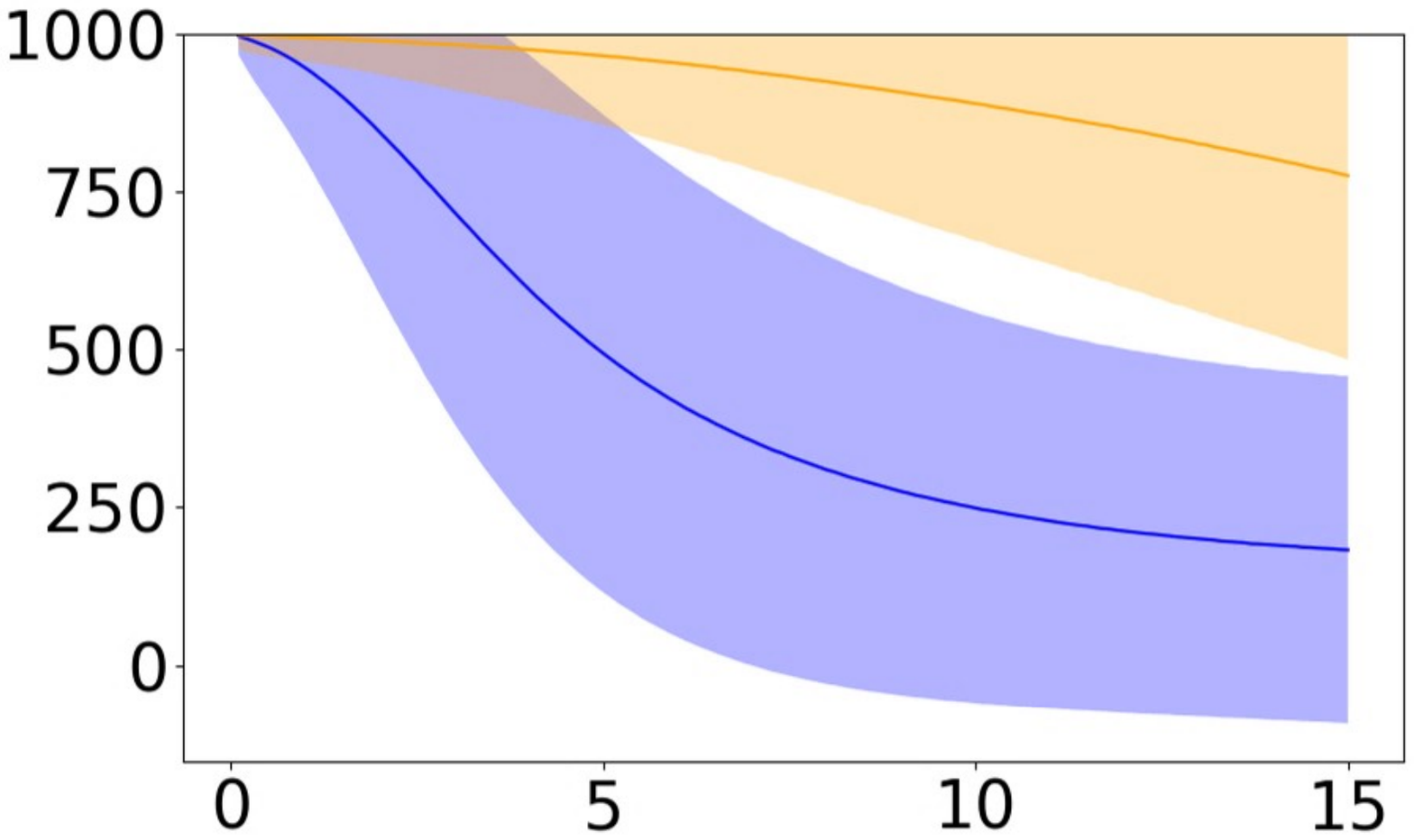}
	}
	\subfigure[Dog, Frog, Horse, Ship, Truck categories, respectively]
	{\label{second_row}
		\includegraphics[trim = 0.5cm 0.5cm 1cm 1cm,clip,width=.19\textwidth]{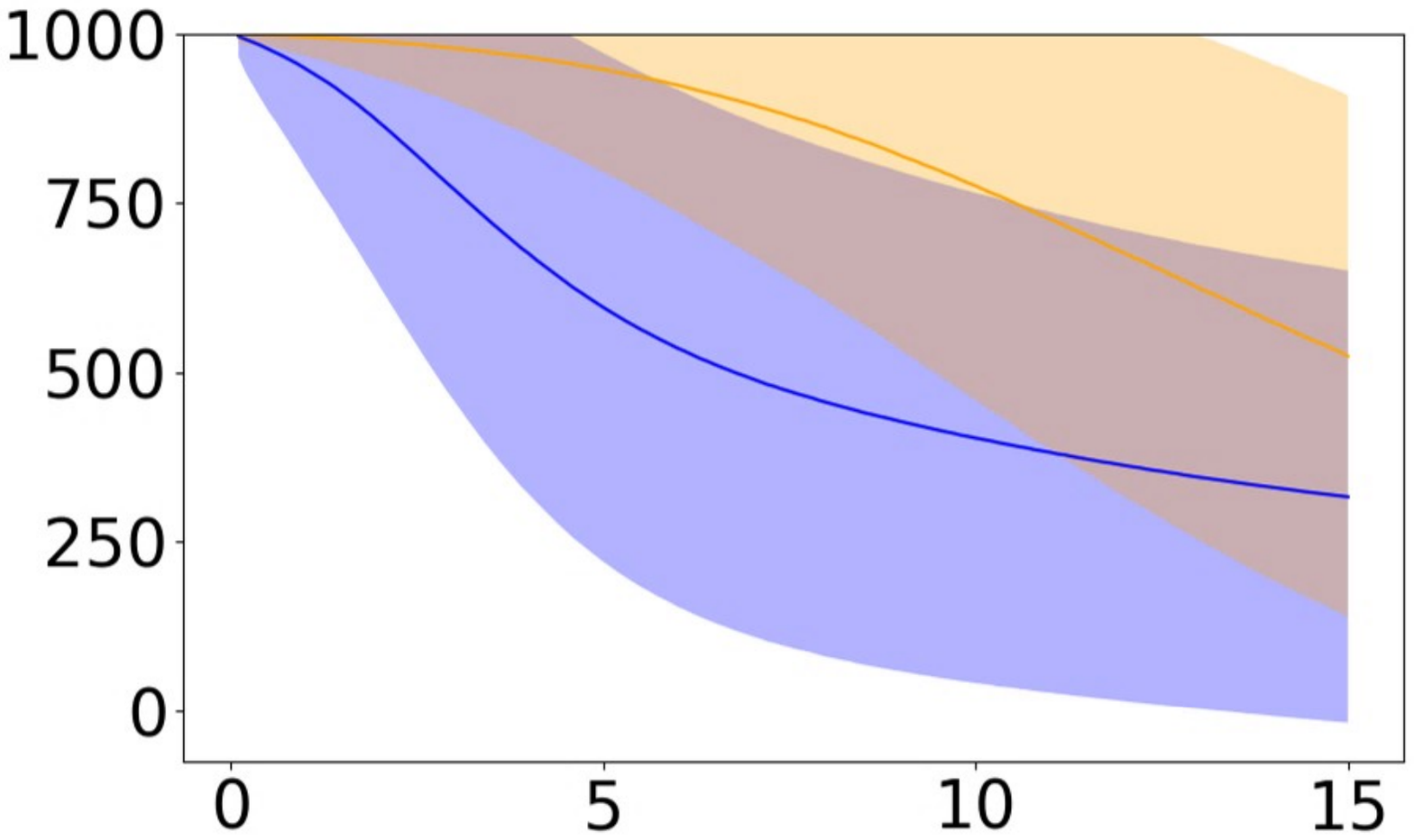}
		\includegraphics[trim = 0.5cm 0.5cm 1cm 1cm,clip,width=.19\textwidth]{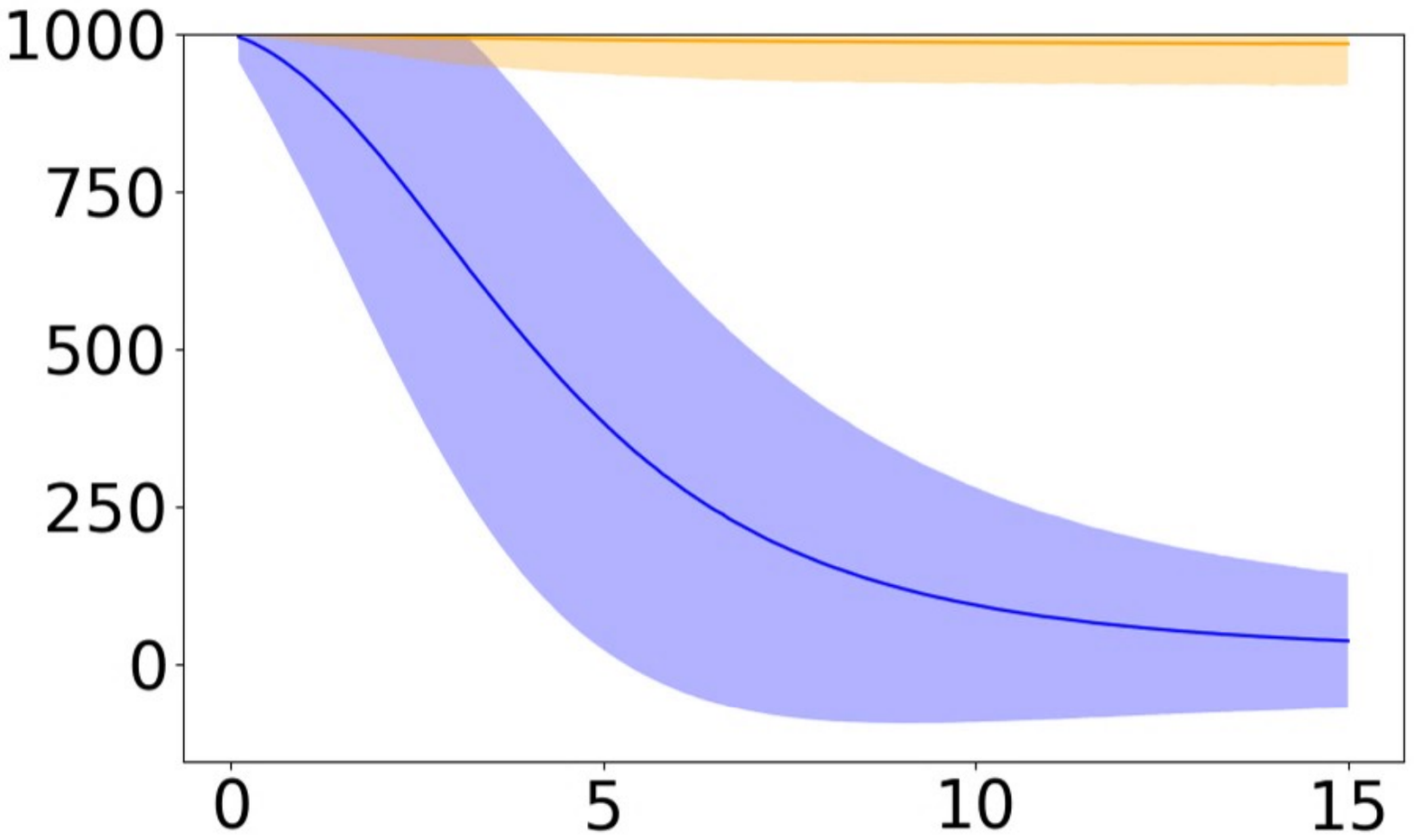}
		\includegraphics[trim = 0.5cm 0.5cm 1cm 1cm,clip,width=.19\textwidth]{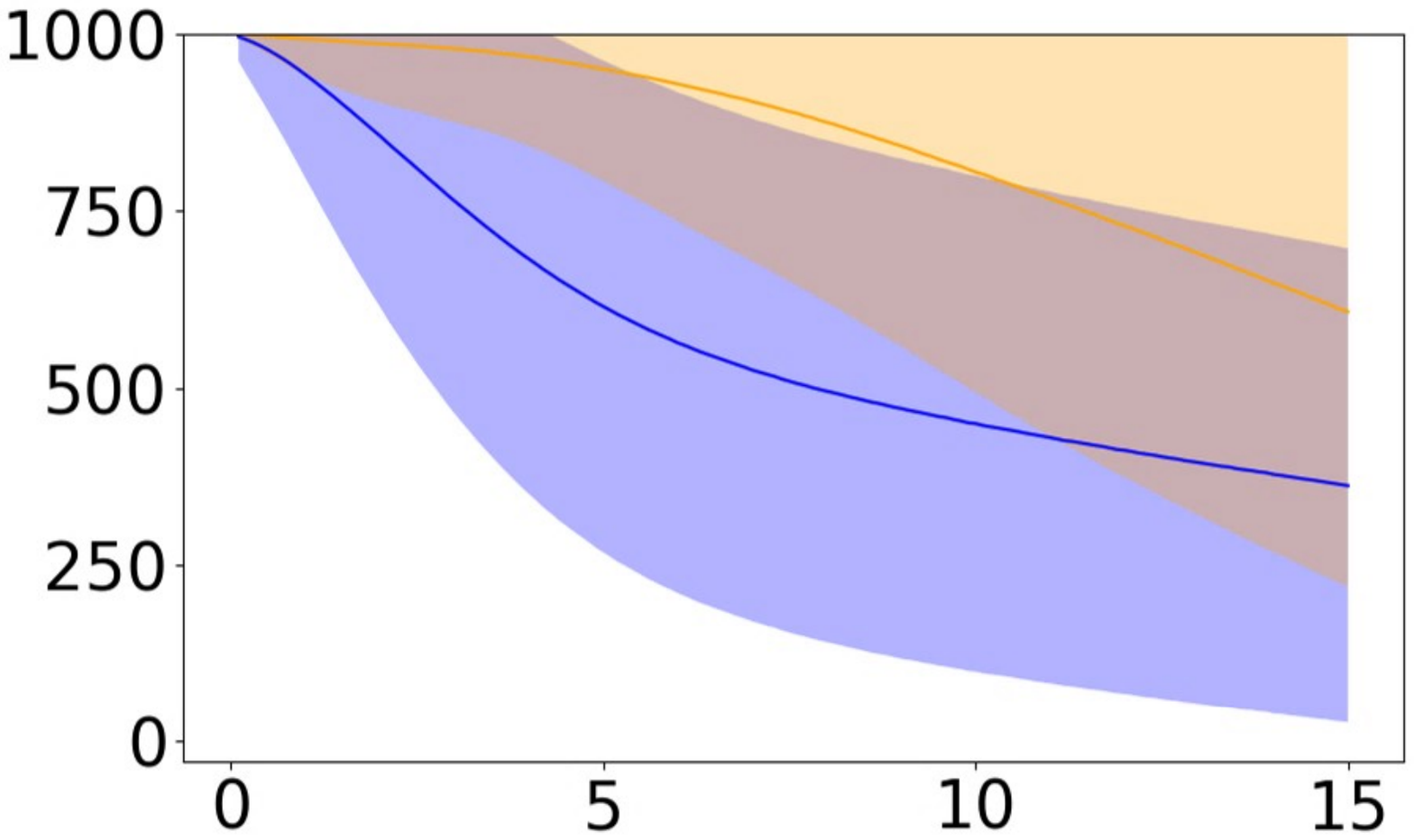}
		\includegraphics[trim = 0.5cm 0.5cm 1cm 1cm,clip,width=.19\textwidth]{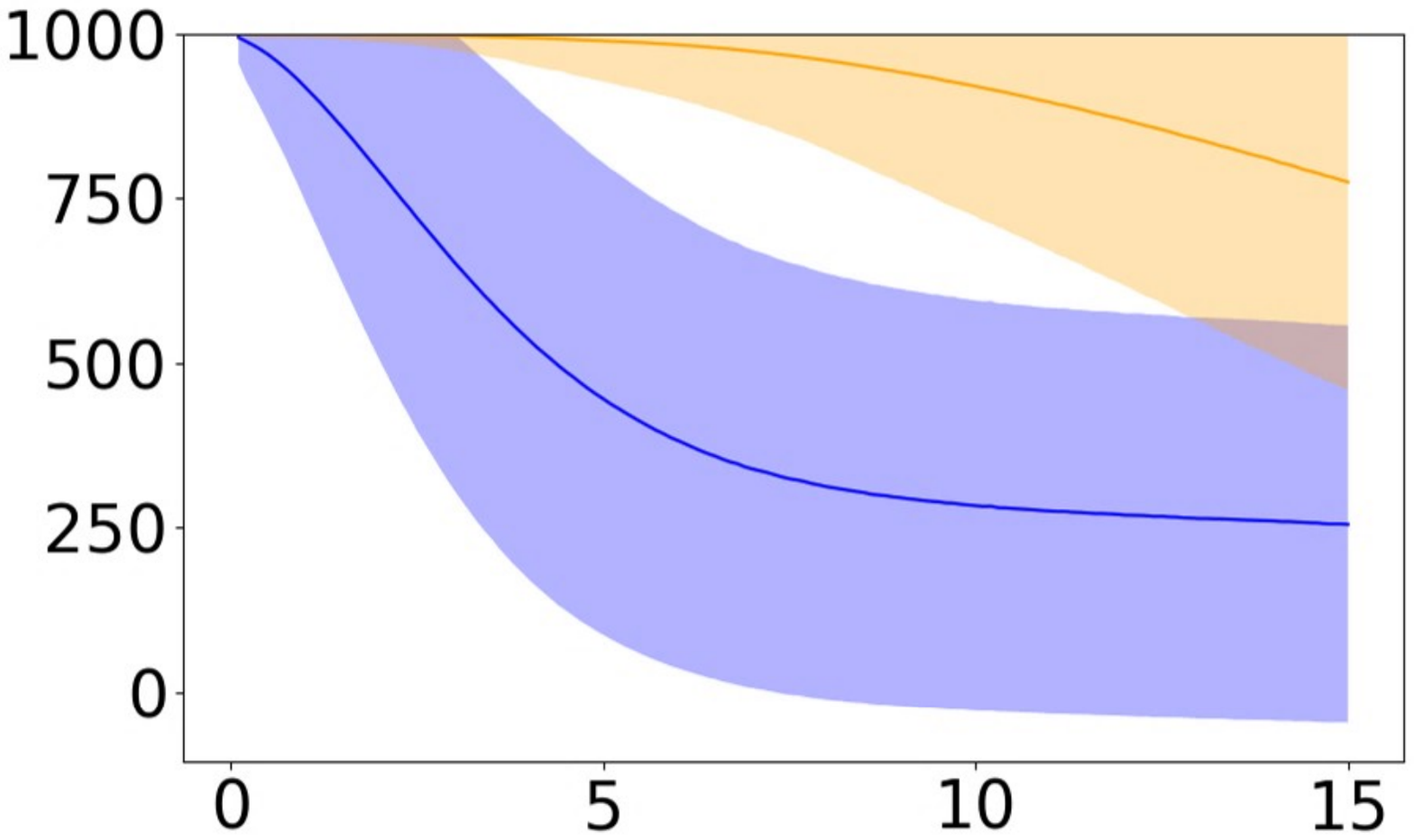}	
		\includegraphics[trim = 0.5cm 0.5cm 1cm 1cm,clip,width=.19\textwidth]{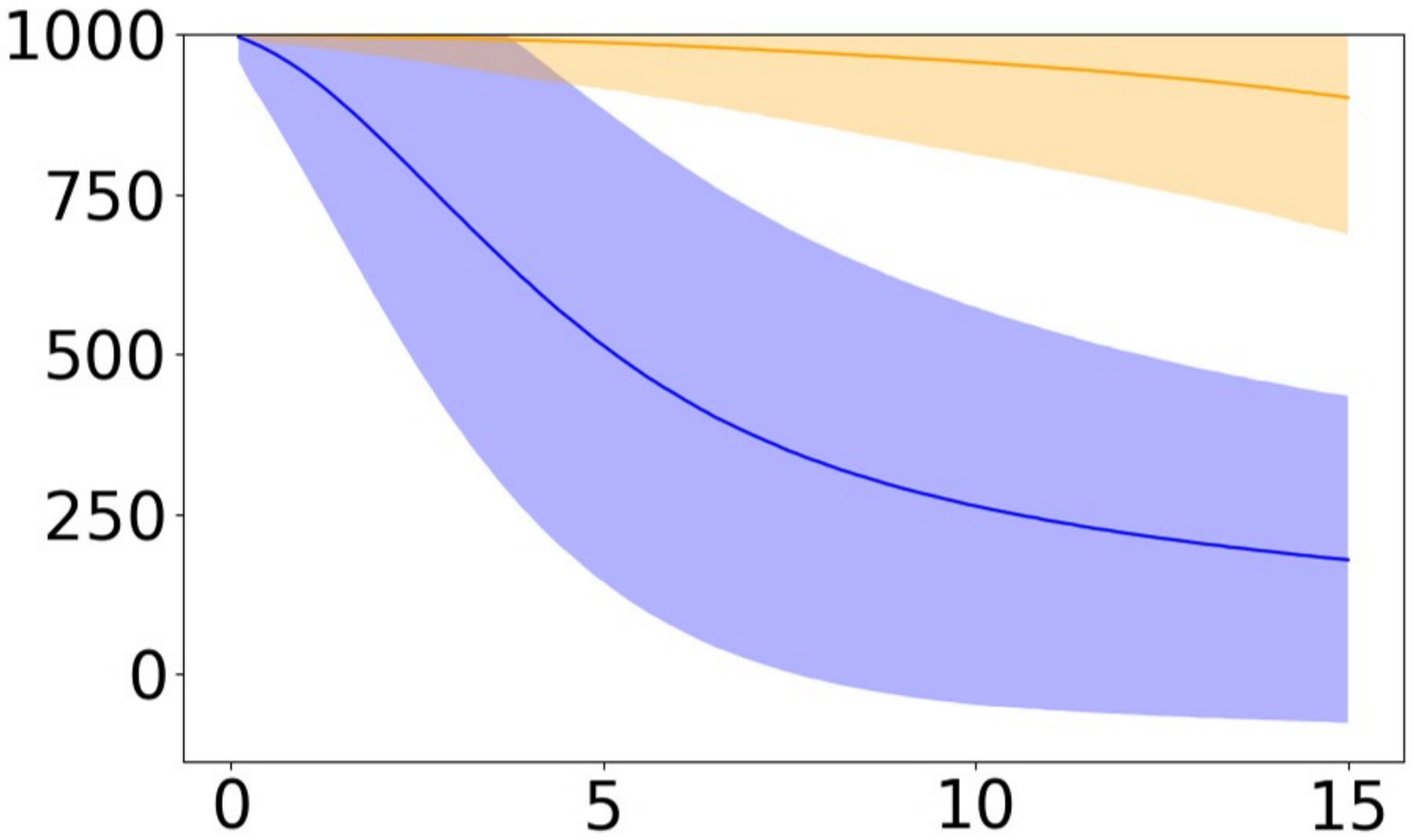}
	}
	\caption{An illustration of CIFAR-10 classifier performance. Each plot corresponds to a single
		output category. The $Y$-axes in the plots show 
		the average number of hits (bold line) and
		standard deviation (faded area), for $k=1000$ perturbations around
		genuine (orange) and adversarial (blue) inputs; whereas the
		$X$-axes represent different values of $\epsilon$ (in percents). The goal
		is to have as small overlap as possible between the two 
		distributions, observing the significance of $\epsilon$ and the variance of the distributions between different categories. 
	}
	\label{fig:Figure1}

\end{figure*}

Our approach requires an offline calibration phase, to adjust the
method's parameters to the DNN at hand. Recall that, during inference,
random samples are procured from an $\epsilon$-region around the input
point $\vec{x}_0$; and that our goal is to select $\epsilon$ in such a
way that many of these samples will be classified the same way as
$\vec{x}_0$ when it is genuine, but only a few will meet this
criterion when $\vec{x}_0$ is adversarial. As it turns out, different
$\epsilon$ values can cause dramatic differences in the number of
hits; see Fig.~\ref{fig:Figure1} for an illustration. In order to
select an appropriate value of $\epsilon$, we use an empirical
approach that leverages Levy et al.'s method~\cite{LeYeKa23} for computing PGCR
values: we test multiple potential $\epsilon$ values, and select
the one that produces the most accurate result.

\mysubsection{Dataset Creation.}  
First, we obtain a set of $n$ correctly classified (genuine) inputs
from the test data, $X_g=\{\vec{x}^g_1,\ldots,\vec{x}^g_n\}$. From
this set, we construct a set $X_a=\{\vec{x}^a_1, \ldots,\vec{x}^a_{n_a}\}$
of adversarial inputs, obtained from the points in $X_g$ by applying
state-of-the-art attacks, such as PGD~\cite{MaMaScTsVl17,MuBrBaLiJo22}. We also consider a set of
$m$ potential $\epsilon$ values,
$E=\{\epsilon_1,\ldots,\epsilon_{m}\}$, from among which we will
select the $\epsilon$ value to be used during inference.

Next, for each $\epsilon\in E$ and for each input point
$\vec{x}\in X=X_g\cup X_a$, we sample $k$ perturbed inputs
$\vec{x}'_1\ldots \vec{x}'_k$ from the $L_{\infty}$ $\epsilon$-region
around $\vec{x}$~\cite{LeKa2021,LeYeKa23}. We then evaluate $N$ on
$\vec{x}$ and on its perturbations $\vec{x}'_1\ldots \vec{x}'_k$,
obtaining the outputs $N(\vec{x})$ for $\vec{x}$ and
$N(\vec{x}'_1)\ldots N(\vec{x}'_k)$ for the perturbations.  We count
the number of \emph{hits},
$h^{\vec{x}}_\epsilon=|\{i\ |\argmmax{N(\vec{x}'_i)}=\argmmax{N(\vec{x})}\}|$,
and store it for later use.

We note that instead of using the measure of hits as a predictor for
whether the input being examined is adversarial, one could try to
discover the distribution of adversarial inputs, and then attempt to
decide whether an input is likely to belong to that
distribution. However, determining the distribution of adversarial
inputs is difficult, and presently cannot be achieved using state-of-the-art
statistical tools~\cite{JMP}.

\subsection{Maximal-Recall-Oriented Calibration}
\label{calibration_recall}
Using the collected data for various $\epsilon$ values, we present
Alg.~\ref{algorithm1} for selecting an optimal 
$\epsilon$ value, and its corresponding threshold $\mathcal{T}$. The
sought-after $\epsilon$ is optimal in the sense that, when used in the
inference phase, it makes the fewest classification mistakes. The
selection algorithm is brute-force: we consider each
$\langle \epsilon, \mathcal{T}\rangle$ pair in turn, compute its success
rate, and then pick the most successful pair.

The algorithm receives as input the set of potential $\epsilon$ values
$E$; the sets of genuine and adversarial points, $X_g$ and $X_a$,
respectively; the measured number of hits, $h^{\vec{x}}_\epsilon$, for
each input point $\vec{x}$ and each $\epsilon$; and also the
hyperparameter $w_g$, discussed later. The algorithm's outputs are
the selected $\epsilon$ value, denoted $\epsilon^*$, and its
corresponding threshold of hits, $\mathcal{T}$, to be used during the
inference phase.

\begin{algorithm}[H]
	\caption{Maximal-Recall-Oriented Calibration}
	\label{algorithm1}
	\hspace*{\algorithmicindent} \textbf{Input:}
	$E$, $X_g$, $X_a$, $\{h^{\vec{x}}_\epsilon\}$, $w_g$ \\
	\hspace*{\algorithmicindent} \textbf{Output:} $\epsilon^{*}, \mathcal{T}$
	\begin{algorithmic}[1]
		\STATE $\epsilon^{*}, bestScore, \mathcal{T} \leftarrow 0,0,0$
		\FORALL{$\epsilon \in E$}\label{alg1:mainLoop}
		\FOR {$t:=0 \ldots k$} \label{alg1:innerLoop}
		\STATE $r_g = |\{ \vec{x}\in X_g \ |\ h_\epsilon^{\vec{x}} > t\}|$ / $|X_g|$
		\label{alg1:computeRecallG}
		\STATE $r_a = |\{ \vec{x}\in X_a \ |\ h_\epsilon^{\vec{x}} <
		t\}|$ / $|X_a|$
		\label{alg1:computeRecallA}
		\STATE {$score = w_g \cdot r_g + (1 - w_g) \cdot r_a$} \label{alg1:computeScore}
		\IF{$score > bestScore$} 
		\STATE {$\epsilon^{*}\leftarrow \epsilon, bestScore\leftarrow
			score, \mathcal{T}\leftarrow t$}\label{alg1:bestScore}
		\ENDIF
		\ENDFOR
		\ENDFOR
		\RETURN {$\epsilon^{*},\mathcal{T}$}\label{alg1:return}
	\end{algorithmic}
\end{algorithm}

Line~\ref{alg1:mainLoop} is the algorithm's main loop,
iterating over all potential $\epsilon$ values in order to pick the
optimal one. For each such $\epsilon$, we iterate over all possible
threshold values in Line~\ref{alg1:innerLoop}; since each pair
$\langle \epsilon,\vec{x}\rangle$ was tested over $k$ perturbed
inputs, the number of hits may vary between $0$ and $k$, and we seek the best choice. Line~\ref{alg1:computeRecallG} calculates the
fraction of genuine inputs for which the threshold was reached,
whereas Line~\ref{alg1:computeRecallA} computes the fraction of
adversarial inputs for which the threshold was not reached; these
values are called the \emph{recall rates}, and are denoted $r_g$ and
$r_a$, respectively. Then, Line~\ref{alg1:computeScore} computes a
real-valued score for this candidate $\epsilon$. The score is a
weighted average between $r_g$ and $r_a$, where the hyperparameter
$w_g$ is used in prioritizing between a more conservative
classification of inputs ($w_g$ is lower, leading to fewer
adversarial inputs mistakenly classified as genuine), or a less
conservative one. The best score, and its corresponding $\epsilon$ and
threshold, are stored in Line~\ref{alg1:bestScore}; and the selected
$\epsilon$ and threshold are returned in Line~\ref{alg1:return}.

\mysubsection{Note.} Alg.~\ref{algorithm1} computes a single
$\epsilon$ value to be used during inference. However, in practice we
used different $\epsilon$ value for each of the different output classes (see
Fig.~\ref{fig:Figure1}). E.g., during inference, if the input at hand
is classified as ``Dog'', one $\epsilon$ value is used; whereas if it is
classified as ``Cat'', another is selected. To make this adjustment,
Alg.~\ref{algorithm1} needs to be run once for each output class, and
the sets $X_a$ and $X_g$ need to be partitioned by output class, as
well. For brevity, we omit these adjustments here, as well as in
subsequent algorithms.

\subsection{Maximal-Precision-Oriented Calibration}
\label{calibration_precision}

\begin{figure}

	\centering
	\subfigure[Recall Calibration Goals]
	{
		\includegraphics[trim =8cm 8.5cm 10cm 5.5cm, clip,width=.19\textwidth]{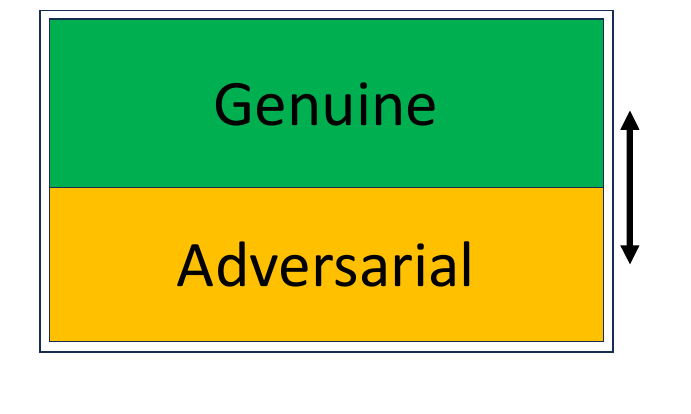}
		\label{fig:calibration_recall_figure}
	}
	\subfigure[Precision Calibration Goals]
	{
		\includegraphics[trim = 8cm 6.5cm 10cm 5.5cm, clip,width=.19\textwidth]{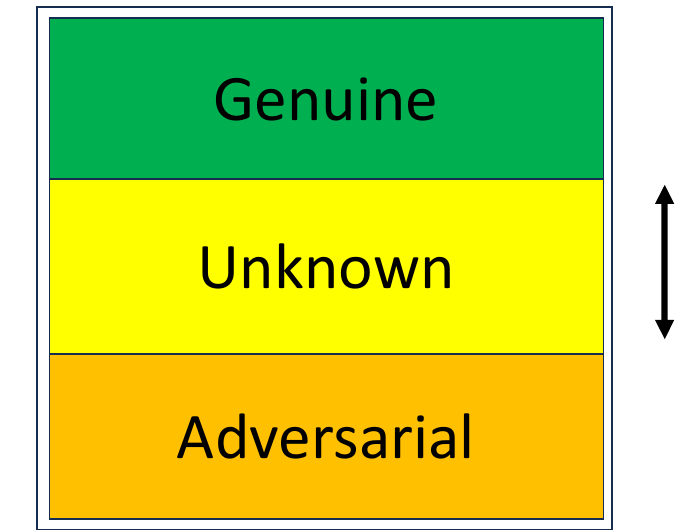}
		\label{fig:precision_calibration_figure}
	}
	\caption{Different optimization	goals: the
		recall-oriented algorithm seeks the optimal threshold value
		that maximizes the separation between genuine and adversarial
		instances. The
		precision-oriented algorithm strives to minimize the
		``yellow'' area, by determining the lowest threshold that
		maximizes adversarial input detection and the highest threshold
		that maximizes genuine input detection, e.g., minimizing the
		``unknown'' cases for both thresholds.}
	\label{fig:precisionVsRecall}
\end{figure}

Alg.~\ref{algorithm1} is geared for recall; i.e., during inference, it
classifies every input as either genuine (if the number of hits
exceeds the threshold), or as adversarial. However, in our
studies we observed that often, a division into \emph{three} classes
was more appropriate: in many cases, inputs encountered during
inference would either demonstrate a very high number of hits, in
which case they are likely genuine; a very low number of hits, in
which case they are likely adversarial; or a medium number of hits, in which
case it is unclear whether they were adversarial or not. While in
Alg.~\ref{algorithm1} a human would have to inspect all inputs
classified as adversarial, a more subtle division into three classes
could allow the human to only inspect those inputs in the
third, ``unknown'' category. See Fig.~\ref{fig:precisionVsRecall} for
an illustration.

To support this variant, we present Alg.~\ref{algorithm2}, which is
geared towards increased \emph{precision}. It is similar to
Alg.~\ref{algorithm1}, but instead of selecting an $\epsilon$ and
returning a single threshold $\mathcal{T}$, it now returns a couple of
thresholds --- $\mathcal{T}_g$, which indicates a lower bar of hits
for genuine inputs, and $\mathcal{T}_a$, which 
indicates an upper bar of
hits for adversarial inputs.  When this algorithm is used, the final
inference step is changed to:
\begin{itemize}
	\item If $h$ is \emph{above} a certain threshold $\mathcal{T}_g$, certify the
	prediction of $N(\vec{x}_0)$ as correct. 
	If $h$ is \emph{below} a certain threshold $\mathcal{T}_a$, mark it as adversarial. 
	Otherwise, flag it as ``unknown''.
\end{itemize}

We observe that this three-output scheme has an inherent tension between precision and recall. For example,
remembering that genuine inputs tend to have a high number of hits and
adversarial inputs a lower number, one could set $\mathcal{T}_g$ to be
very high and $\mathcal{T}_a$ to be very low. This would result in
excellent precision (i.e., the algorithm would make only a few
mistakes), but many inputs would fall into the ``unknown'' category,
causing poor recall. In terms of Fig.~\ref{fig:precisionVsRecall},
this would result in a very wide yellow strip. Alternatively, one
could set these two thresholds to be near-identical (resulting in a
narrow yellow strip), which would simply mimic
Alg.~\ref{algorithm1}, with its excellent recall but sub-optimal
precision. To resolve this, we allow the user to specify a-priori the
desired precision, by setting minimal values $p_g^{\min}$ and
$p_a^{\min}$ for the precision over genuine and adversarial inputs,
respectively. Our algorithm then automatically selects the $\epsilon$
and threshold values that meet these requirements, and which achieve
the maximum recall among all candidates.

More concretely, the inputs to Alg.~\ref{algorithm2} are the same as
those to Alg.~\ref{algorithm1}, plus the two threshold parameters,
$p_g^{\min}$ and $p_a^{\min}$.  For each candidate $\epsilon$, the
algorithm iterates over all possible thresholds
(Line~\ref{alg2:innerLoop1}), from high to low, each time computing the
recall rates (Lines~\ref{alg2:rg1}--\ref{alg2:ra1}) and the precision
rate over genuine inputs (Line~\ref{alg2:pg}). If the precision rate
meets the specified threshold, the threshold is stored as $t_1$  in
Line~\ref{alg2:storeT1}. At the end of this loop, the algorithm has
identified the lowest threshold for genuine inputs that meet the
requirements.  Next, the algorithm begins a symmetrical process for
the threshold for adversarial inputs, this time iterating from low to
high (Line~\ref{alg2:innerLoop2}), until it identifies the greatest
threshold that satisfies the requirements ($t_2$, Line~\ref{alg2:storeT2}).
Once both thresholds are discovered, the algorithm uses the recall
ratios to compute a score (Line~\ref{alg2:computeScore}) --- in order
to compare the various $\epsilon$ values. The most optimal $\epsilon$ and its
thresholds are stored in Line~\ref{alg2:bestScore}, and are finally
returned in Line~\ref{alg2:return}.

\begin{algorithm}[ht]
	\caption{Maximal-Precision-Oriented Calibration}
	\label{algorithm2}
	\hspace*{\algorithmicindent} \textbf{Input:}
	$E$, $X_g$, $X_a$, $\{h^{\vec{x}}_\epsilon\}$, $w_g$,
	$p_g^{\min}$, $p_a^{\min}$ \\
	\hspace*{\algorithmicindent} \textbf{Output:} $\epsilon^{*}$, $\mathcal{T}_g$,$\mathcal{T}_a$
	\begin{algorithmic}[1]
		
		\STATE $\epsilon^{*},bestScore, \mathcal{T}_g,\mathcal{T}_a,r_{\mathcal{T}_g},r_{\mathcal{T}_a} \leftarrow 0,0,0,0,0,0$
		\FORALL{$\epsilon \in E$}\label{alg2:mainLoop}
		\STATE {$r_1,r_2\leftarrow 0,0$}
		\FOR {$t:=k \ldots 0$} \label{alg2:innerLoop1}
		\STATE $r_g = |\{ \vec{x}\in X_g \ |\ h_\epsilon^{\vec{x}} > t\}|$ \label{alg2:rg1}
		/ $|X_g|$
		\STATE $r_a = |\{ \vec{x}\in X_a \ |\ h_\epsilon^{\vec{x}} < t\}|$ \label{alg2:ra1}
		/ $|X_a|$
		\STATE {$p_g = r_g / \left(r_g+(1-r_a)\right)$} \label{alg2:pg}
		\IF {$p_g > p_g^{\min}$}
		\STATE $t_1 \leftarrow t$, $r_1\leftarrow r_g$ \label{alg2:storeT1}
		\ELSE
		\STATE break
		\ENDIF
		\ENDFOR
		
		\FOR {$t:=1 \ldots t_1$} \label{alg2:innerLoop2}
		\STATE $r_g = |\{ \vec{x}\in X_g \ |\ h_\epsilon^{\vec{x}} > t\}|$
		/ $|X_g|$
		\STATE $r_a = |\{ \vec{x}\in X_a \ |\ h_\epsilon^{\vec{x}} < t\}|$
		/ $|X_a|$
		\STATE {$p_a = r_a / \left(r_a+(1-r_g)\right)$}
		\IF {$p_a > p_a^{\min}$}
		\STATE {$t_2 \leftarrow t$, $r_2\leftarrow r_a$} \label{alg2:storeT2}
		\ELSE
		\STATE break
		\ENDIF
		\ENDFOR
		\STATE $score=w_g \cdot r_1 + (1-w_g)\cdot r_2$ \label{alg2:computeScore}
		\IF{$(score > bestScore)$}  \label{alg2:bestScore}
		\STATE {$\epsilon^{*}\leftarrow \epsilon, bestScore\leftarrow
			score, \mathcal{T}_g \leftarrow t_1,\mathcal{T}_a \leftarrow t_2 $}
		\ENDIF
		\ENDFOR
		\RETURN $\epsilon^{*}, \mathcal{T}_g,\mathcal{T}_a$\label{alg2:return}
	\end{algorithmic}
\end{algorithm}

\section{Evaluation}
\label{Evaluation}

\mysubsection{Implementation.} 
In order to evaluate our method, we
created a proof-of-concept implementation of Algs.~\ref{algorithm1}
and~\ref{algorithm2} in a tool called DEM, which is available
online~\cite{ourCode}.
Our tool is written in Python, and supports DNNs
in the common PyTorch format.

\mysubsection{Baseline.}
In our experiments we compared the two
algorithms within DEM (recall-oriented and precision-oriented), and
also compared DEM's recall-oriented algorithm to the Local Intrinsic
Dimension (LID) method~\cite{MaLiWaErWiScSoHoBa18}, which is the
state of the art in detecting adversarial examples. Like \ourTool, LID
has a calibration phase, in which the network is evaluated on
multiple genuine and adversarial inputs. LID uses these evaluations to
examine the assignments to the various neurons of the DNN, and then trains
a regression model to predict, during inference and based on these values,
whether an input is adversarial.

Since LID was not originally designed for
categorical data, we trained it using two separate methods:
\begin{inparaenum}[(i)]
\item using examples from all classes, as originally
suggested~\cite{MaLiWaErWiScSoHoBa18}; and
\item training ten LID detectors, one for each output class.
\end{inparaenum}
The second approach improved LID's performance slightly, and we used
it as benchmark in the experiments described next.

\mysubsection{Benchmarks.} 
We used VGG16~\cite{SiZi15} and
Resnet~\cite{HeZhReSu16} DNN models, trained on the CIFAR10
dataset~\cite{KrHi09}. The Resnet and VGG models achieved $88.35\%$
and $83.75\%$ accuracy, respectively. We then took the CIFAR10 test
set, and removed from it any inputs that were misclassified by the trained
models. We split the remaining inputs into two sets: a calibration
set with $80\%$ of the inputs, and an evaluation set with the
remaining $20\%$.  The adversarial inputs for the calibration and
evaluation phases were obtained using PGD~\cite{MaMaScTsVl17}, with a
maximal modification distance of $0.005$ (i.e., $0.5\%$).

In the preliminary calibration phase, we created $1000$ perturbed
inputs around each genuine and adversarial input. We empirically
observed that $1000$ samples was a suitable choice ---
adding additional samples did not change our algorithm's performance
significantly, whereas using significantly fewer samples led to degraded performance. 
For both calibration methods, we set the
genuine recall hyperparameter to $w_g=0.3$.
This conservative choice
prioritizes avoiding adversarial inputs wrongly misclassified as
genuine.  For the set $E$ of candidate $\epsilon$ values, we
arbitrarily chose 
distance values ranging from $0.001$ to $0.15$, with $0.001$
resolution. While RoMA and gRoMA used an $\epsilon$ value of
$0.04$~\cite{LeKa2021,LeYeKa23}, we decided to use a higher value to ensure we do not miss a better calibration value.

We note that large values of $\epsilon$ reduce the number of
hits for genuine inputs, because the region includes legitimate inputs
with true different classifications; we empirically found
that a distance greater than $0.15$ typically causes this number to
drop steeply.

We conducted our evaluation on a standard laptop, equipped with an AMD
Ryzen 9 6900HX CPU, an NVIDIA GeForce RTX 3070 GPU, and 16GB of
RAM. In order to create a virtual running environment, we used Python
3.11 and PyTorch 2.11.  
The calibration and data preparation for each algorithm took less than four hours per class. Each sample was analyzed in less than 0.003 seconds.
The code for the creation of adversarial examples, calibration, and evaluation is available online~\cite{ourCode}.

\subsection{Evaluating the Recall-Oriented Calibration Algorithm}

Table~\ref{table:recall_results} provides a comparison of our
recall-oriented calibration algorithm with
LID~\cite{MaLiWaErWiScSoHoBa18}. It demonstrates that the proposed
method is highly effective at distinguishing between genuine and
adversarial inputs. Compared to LID, we observe that LID identifies
genuine examples with a slightly better recall; however, \ourTool{} is
significantly better at identifying adversarial inputs for all classes. 
This is a favorable result, as the mistake of
classifying an adversarial input as a genuine one is considered to be
much more serious, in the use-cases we consider, than the second kind of mistake --- classifying a
genuine input as an
adversarial.

\begin{table}[ht]
	\centering
	\caption{Evaluating the recall-oriented calibration algorithm, in comparison to LID. The evaluation includes the Resnet (marked as \textcolor{red}{RES}) and VGG (marked as \textcolor{blue}{VGG}) models. The two parts of the table present genuine and adversarial recall: the number of samples, the recall achieved by \ourTool, and the recall achieved by LID.}
	\label{table:recall_results}
	\setlength{\tabcolsep}{3pt} 
	\renewcommand{\arraystretch}{0.8} 
	\scriptsize 

	\begin{tblr}{
			hline{1,3,Z}=1pt, hline{2}={solid}, hline{4-Y} ={dashed}, vlines,
			colspec = {l *{6}{X[0.75,l]}}, 
			rowsep = 2pt 
		}
		\SetCell[r=2]{l} Class
		& \SetCell[c=3]{c} Genuine Recall
		& & & \SetCell[c=3]{c} Adversarial Recall
		& & \\
		& \#Samples \textcolor{red}{RES}, \textcolor{blue}{VGG} & \ourTool & LID
		& \#Samples \textcolor{red}{RES}, \textcolor{blue}{VGG} & \ourTool & LID\\
		\hline
		Airplane & \textcolor{red}{180},\textcolor{blue}{177} & \textcolor{red}{63\%},\textcolor{blue}{76\%} & \textcolor{red}{81\%},\textcolor{blue}{88\%} & \textcolor{red}{393},\textcolor{blue}{125} & \textcolor{red}{92\%},\textcolor{blue}{86\%} & \textcolor{red}{54\%},\textcolor{blue}{30\%} \\
		Automotive & \textcolor{red}{190},\textcolor{blue}{185} & \textcolor{red}{65\%},\textcolor{blue}{65\%} & \textcolor{red}{80\%},\textcolor{blue}{91\%} & \textcolor{red}{290},\textcolor{blue}{80} & \textcolor{red}{94\%},\textcolor{blue}{87\%} & \textcolor{red}{43\%},\textcolor{blue}{52\%} \\
		Bird & \textcolor{red}{169},\textcolor{blue}{160} & \textcolor{red}{79\%},\textcolor{blue}{67\%} & \textcolor{red}{80\%},\textcolor{blue}{84\%} & \textcolor{red}{441},\textcolor{blue}{159} & \textcolor{red}{91\%},\textcolor{blue}{91\%} & \textcolor{red}{44\%},\textcolor{blue}{35\%} \\
		Cat & \textcolor{red}{149},\textcolor{blue}{146} & \textcolor{red}{51\%},\textcolor{blue}{50\%} & \textcolor{red}{74\%},\textcolor{blue}{85\%} & \textcolor{red}{547},\textcolor{blue}{246} & \textcolor{red}{91\%},\textcolor{blue}{89\%} & \textcolor{red}{52\%},\textcolor{blue}{31\%} \\
		Deer & \textcolor{red}{179},\textcolor{blue}{176} & \textcolor{red}{80\%},\textcolor{blue}{79\%} & \textcolor{red}{76\%},\textcolor{blue}{89\%} & \textcolor{red}{526},\textcolor{blue}{186} & \textcolor{red}{93\%},\textcolor{blue}{97\%} & \textcolor{red}{47\%},\textcolor{blue}{37\%} \\
		Dog & \textcolor{red}{164},\textcolor{blue}{135} & \textcolor{red}{56\%},\textcolor{blue}{60\%} & \textcolor{red}{74\%},\textcolor{blue}{87\%} & \textcolor{red}{465},\textcolor{blue}{183} & \textcolor{red}{91\%},\textcolor{blue}{92\%} & \textcolor{red}{44\%},\textcolor{blue}{30\%} \\
		Frog & \textcolor{red}{183},\textcolor{blue}{174} & \textcolor{red}{99\%},\textcolor{blue}{91\%} & \textcolor{red}{76\%},\textcolor{blue}{84\%} & \textcolor{red}{452},\textcolor{blue}{138} & \textcolor{red}{99\%},\textcolor{blue}{97\%} & \textcolor{red}{54\%},\textcolor{blue}{34\%} \\
		Horse & \textcolor{red}{183},\textcolor{blue}{174} & \textcolor{red}{65\%},\textcolor{blue}{74\%} & \textcolor{red}{84\%},\textcolor{blue}{88\%} & \textcolor{red}{404},\textcolor{blue}{125} & \textcolor{red}{88\%},\textcolor{blue}{87\%} & \textcolor{red}{44\%},\textcolor{blue}{38\%} \\
		Ship & \textcolor{red}{188},\textcolor{blue}{185} & \textcolor{red}{88\%},\textcolor{blue}{91\%} & \textcolor{red}{79\%},\textcolor{blue}{92\%} & \textcolor{red}{260},\textcolor{blue}{64} & \textcolor{red}{96\%},\textcolor{blue}{92\%} & \textcolor{red}{57\%},\textcolor{blue}{56\%} \\
		Truck & \textcolor{red}{185},\textcolor{blue}{184} & \textcolor{red}{86\%},\textcolor{blue}{80\%} & \textcolor{red}{83\%},\textcolor{blue}{87\%} & \textcolor{red}{331},\textcolor{blue}{100} & \textcolor{red}{95\%},\textcolor{blue}{95\%} & \textcolor{red}{45\%},\textcolor{blue}{44\%} \\
	\end{tblr}
\end{table}

Examining the results, we observe the notable effectiveness of
\ourTool{} when applied to the specific classes of the Resnet
model. For instance, note the Frog class, where detection rates
consistently surpass 90\% across numerous samples. Nevertheless, some
of the other classes demonstrate lesser effectiveness, and
consequently these results, as a whole, do not meet the
aerospace certification standards.

\subsection{Evaluating the Precision-Oriented Calibration Algorithm}

We conducted another similar experiment, this time using
precision-oriented calibration, with precision thresholds set to
$85\%$ for genuine examples on Resnet and VGG,
and $80\%$ for adversarial inputs with both models.
The $w_g$ hyperparameter remained the same as in the recall
evaluation. Table~\ref{table:precision_results} depicts the
results. The recall results reported therein appear alongside their
corresponding precision results, for the same $\epsilon$ values.
Except for one case, all recall values were above $50\%$, and most of them were over $70\%$.
That second algorithm produces a detector with the required precision value in most cases.

\begin{table}[htb]
	\centering
	\caption{Evaluation of the precision-oriented calibration algorithm. Resnet results are marked as \textcolor{red}{RES}, and VGG results as \textcolor{blue}{VGG}.}
	\label{table:precision_results}
	\setlength{\tabcolsep}{3pt} 
	\renewcommand{\arraystretch}{0.8} 
	\scriptsize 
	\begin{tblr}{
			hline{1,3,Z}=1pt, hline{2}={solid}, hline{4-Y} ={dashed}, vlines,
			colspec = {l *{6}{X[0.75,l]}}, 
			rowsep = 2pt 
		}
		\SetCell[r=2]{l} Class
		& \SetCell[c=3]{c} Genuine Examples
		& & & \SetCell[c=3]{c} Adversarial Inputs
		& & \\
		& \#Samples \textcolor{red}{RES}, \textcolor{blue}{VGG} & Precision & Recall
		& \#Samples \textcolor{red}{RES}, \textcolor{blue}{VGG} & Precision & Recall \\
		\hline
		Airplane & \textcolor{red}{180},\textcolor{blue}{177} & \textcolor{red}{88\%},\textcolor{blue}{85\%} & \textcolor{red}{61\%},\textcolor{blue}{80\%} & \textcolor{red}{393},\textcolor{blue}{125} & \textcolor{red}{80\%},\textcolor{blue}{81\%} & \textcolor{red}{68\%},\textcolor{blue}{86\%} \\
		Automotive & \textcolor{red}{190},\textcolor{blue}{185} & \textcolor{red}{91\%},\textcolor{blue}{82\%} & \textcolor{red}{68\%},\textcolor{blue}{73\%} & \textcolor{red}{290},\textcolor{blue}{80} & \textcolor{red}{77\%},\textcolor{blue}{76\%} & \textcolor{red}{91\%},\textcolor{blue}{84\%} \\
		Bird & \textcolor{red}{169},\textcolor{blue}{160} & \textcolor{red}{89\%},\textcolor{blue}{87\%} & \textcolor{red}{79\%},\textcolor{blue}{78\%} & \textcolor{red}{441},\textcolor{blue}{159} & \textcolor{red}{81\%},\textcolor{blue}{81\%} & \textcolor{red}{91\%},\textcolor{blue}{79\%} \\
		Cat & \textcolor{red}{149},\textcolor{blue}{146} & \textcolor{red}{91\%},\textcolor{blue}{83\%} & \textcolor{red}{41\%},\textcolor{blue}{53\%} & \textcolor{red}{547},\textcolor{blue}{246} & \textcolor{red}{77\%},\textcolor{blue}{73\%} & \textcolor{red}{69\%},\textcolor{blue}{55\%} \\
		Deer & \textcolor{red}{179},\textcolor{blue}{176} & \textcolor{red}{90\%},\textcolor{blue}{85\%} & \textcolor{red}{82\%},\textcolor{blue}{91\%} & \textcolor{red}{526},\textcolor{blue}{186} & \textcolor{red}{84\%},\textcolor{blue}{90\%} & \textcolor{red}{91\%},\textcolor{blue}{84\%} \\
		Dog & \textcolor{red}{164},\textcolor{blue}{135} & \textcolor{red}{88\%},\textcolor{blue}{80\%} & \textcolor{red}{50\%},\textcolor{blue}{68\%} & \textcolor{red}{465},\textcolor{blue}{183} & \textcolor{red}{74\%},\textcolor{blue}{75\%} & \textcolor{red}{74\%},\textcolor{blue}{53\%} \\
		Frog & \textcolor{red}{183},\textcolor{blue}{174} & \textcolor{red}{88\%},\textcolor{blue}{83\%} & \textcolor{red}{100\%},\textcolor{blue}{99\%} & \textcolor{red}{452},\textcolor{blue}{138} & \textcolor{red}{100\%},\textcolor{blue}{99\%} & \textcolor{red}{86\%},\textcolor{blue}{80\%} \\
		Horse & \textcolor{red}{183},\textcolor{blue}{174} & \textcolor{red}{85\%},\textcolor{blue}{84\%} & \textcolor{red}{64\%},\textcolor{blue}{79\%} & \textcolor{red}{404},\textcolor{blue}{125} & \textcolor{red}{75\%},\textcolor{blue}{80\%} & \textcolor{red}{80\%},\textcolor{blue}{85\%} \\
		Ship & \textcolor{red}{188},\textcolor{blue}{185} & \textcolor{red}{90\%},\textcolor{blue}{84\%} & \textcolor{red}{95\%},\textcolor{blue}{94\%} & \textcolor{red}{260},\textcolor{blue}{64} & \textcolor{red}{94\%},\textcolor{blue}{93\%} & \textcolor{red}{90\%},\textcolor{blue}{83\%} \\
		Truck & \textcolor{red}{185},\textcolor{blue}{184} & \textcolor{red}{89\%},\textcolor{blue}{87\%} & \textcolor{red}{91\%},\textcolor{blue}{91\%} & \textcolor{red}{331},\textcolor{blue}{100} & \textcolor{red}{91\%},\textcolor{blue}{91\%} & \textcolor{red}{88\%},\textcolor{blue}{86\%} \\
	\end{tblr}
\end{table}

\label{betterAlgorithm} 
Table~\ref{table:precision_results} highlights the fact that taking
into account both Precision and Recall affords a more favorable analysis of sample
robustness.

\section{Discussion and Future Work}
\label{Discussion}

We presented here the \ourTool{} tool, which uses  
a novel, efficient probabilistic certification approach to determine the
trustworthiness of individual DNN predictions.
The proposed technique has several
advantages:
\begin{inparaenum}[(i)]
	\item it is applicable to black-box DNNs;
	\item it is computationally efficient during inference, using
	pre-calibrated parameters; and
	\item our evaluation results highlight \ourTool{}'s
	success  in identifying adversarially distorted inputs.
\end{inparaenum}
\ourTool{} could enable the selective use of DNN outputs when they are
safe, while requesting human oversight when needed.

Moving forward,
we plan to extend the study to cover more specific aviation use cases and related DNNs.
In the long run, we hope that \ourTool{}  will aid in creating
certified  DNN co-pilots. We further hope that including the \ourTool{} variance
profile within functional hazard analyses may facilitate regulatory 
acceptance of classifier DNNs, allowing us to realize more of the AI potential in the aerospace domain.

\bibliographystyle{IEEEtran}
\bibliography{CertifyDnnResults}
\end{document}